\documentstyle[12pt]{article}
\vspace{5cm}
\title{Gluon Distributions in Real and \\
Virtual Photon and the Photon Structure Function}
\vspace{2cm}
\author{B.L.Ioffe\\
Institute of Theoretical and Experimental Physics\\
B.Cheremushkinskaya 25, 117259, Moscow, Russia\\
(e--mail: ioffe@vxitep.itep.ru)\\
FAX 7(095) 883 96 05\\
and\\
A.Oganesian\\
Yerevan Physics Institute\\
Alikhanian Brothers St.,2, 375036, Yerevan, Armenia\\
(e--mail: armen@vxitep.itep.ru)}
\date{}
\newcommand{\be}{\begin{equation}}
\newcommand{\ee}{\end{equation}}
\begin{document}
\maketitle

\large
\vspace{10mm}
\centerline{{\bf Abstract}}

\vspace{5mm}

Gluon distributions in real and virtual photons are calculated using
evolution equations in the NLO approximation. The quark distributions in the
photon determined on the basis of the QCD sum rule approach in ref.[1] are
taken as an input. It is shown that gluon distribution in the photon can
be reliably determined up to $x = 0.03 \div 0.05$ much  lower than the
corresponding values in the case of quark distributions.Two variants of the
calculations are considered:  (1) it is assumed that there are no intrinsic
gluons in the photon at some low normalization point $Q^2 = Q^2_0 \sim 1
GeV^2$; (2) it is assumed that gluonic content of the photon at low $Q^2_0$
is described by gluonic content of vector mesons $\rho, \omega, \varphi$.
The gluon distributions in these two variants appear to be different. This
fact permits one to clarify the origin of nonperturbative gluonic content of
the photon by comparing the results with experiment.

Structure functions $F_2(x)$ for real and virual photon are
calculated and it is shown that in the region $x \geq 0.2$ where QCD
approach is valid, there is a good agreement with experiment.

\newpage
{\bf 1.~ \underline{Introduction}}

\vspace{5mm}
In paper [1] the structure functions of real and virtual photons were
calculated. The hadronic part of the photon structure function was
calculated in a modelless way on the basis of the  QCD sum rules
applying to structure functions [2,3]. In this aspect the calculation
made in [1] differs from earlier  considerations of the  same problem
[4--7]. The idea of the ref.[1]  consideration was the  following. First, the
structure function of the virtual photon (photon--target) was studied under
conditions when photon virtuality $p^2 < 0 $ and $\mid p^2 \mid >>
R^{-2}_c,~ R_c$ -- is the confiniment radius, but at the same time $\mid p^2
\mid << Q^2=-q^2$,  where $q$ -- is the photon projectile momentum. In
expansion in power in $p^2/q^2$ only terms of the first nonvanishing order
were taken into account -- the terms of the lowest twist--2. The operator
product expansion (OPE)  over $1/p^2$ was performed and the first and second
order terms were taken into account. The contribution of the latter
corresponds to interaction of quarks with vacuum gluonic fields and leads to
appearance of terms proportional to gluonic
condensate in the structure function of the virtual photon. Thus, the
structure function of the virtual photon $F_2(s,p^2),~s=(p+q)^2$  was
constructed as a series in $1/p^2$. On the other hand, using analytical
properties of $F_2(s,p^2)$ as functions of $p^2$ by means of dispersion
relations in $p^2$ at fixed  $s~ F_2(s,p^2)$ was presented through
contributions of physical states modelled as contributions of vector mesons
$(\rho, \varphi, \omega)$ and continuum. By identifying both of these
representations  the structure function of the virtual photon
including its hadronic part was found. At such approach the structure
function possesses correct analytical properties in    $p^2$  and has no
fictitious singularties at $p^2=0$ inherent to bare diagram of Fig.1 (for
massless quarks).

The result of the calcultions of the transverse photon structure function is
[1]:

$$(\frac{3\alpha}{\pi})^{-1}~F^T_2(x,p^2,Q^2)=x~\sum_{q}e^4_q~\{~[2+\kappa(x)(ln~
\frac{Q^2}{x^2(s_0-p^2)}-3+$$
\be
+\frac{p^2}{s_0-p^2})~] +
\frac{1}{2}~\frac{s^2_0}{(p^2-m^2_{V,q})^2}~[\kappa(x)
-\frac{8\pi^2}{27 s^2_0 x^2}< 0\mid \frac{\alpha_s}{\pi}G^2_{\mu\nu}\mid
0>]~ \},
\ee
where $x=Q^2/2\nu$ is  the Bjorken variable, $\kappa(x)=x^2+(1-x)^2,~ s_0=1.5
GeV^2$ is the  continuum threshold, $m_{V,q}$ are the vector meson masses,
$\rho$ and $\omega$ for $q=u, d$ and $\varphi$ for $q=s$. (We restrict
ourselves to consideration of three flavours).

As has been shown in [1] and as is seen from (1) the first term in square
brackets in the right--hand side of (1) corresponds to the continuum
contribution in dispersion representation in $p^2$ while the second to the
vector-meson contribution. Applying duality considerations we arrive at the
conclusion that the first term in the right--hand side of (1) can
be considered as a photon perturbative contribution into $F_2$,
and the second term as hadronic part of the  structure function $F_2$.
Thereby, we unambiquously select from the diagram of Fig.1 the soft and
collinear quark contribution which can referred to quark sea distribution in
photon but not to direct contribution of photonic operators in OPE.

At $x> 0.2$ the gluonic condensate contribution in (1) is less than 15$\%$
even at $p^2=0$. This allows one (at these values of $x$) to extrapolate (1)
to the point $p^2=0$ and to get structure function of the real photon.

An essential disadvantage of the discussed method of the structure functions
calculation
[2,3,1] is that it is inapplicable at small $x$ and $x$
close to 1. As is seen from (1), the correction term of OPE proportional to
gluonic condensate fastly increases with $x$ decreasing, i.e. the series in
$1/p^2$ diverges at small $x$. This is  caused by the fact that in
the imaginary part of the forward $\gamma-\gamma$ scattering amplitude --
the diagrams of Fig.1 -- the quark virtuality at vertical quark lines  $k^2
\sim xp^2$ for massless quarks. Thus, at small $x$ quarks appear to be in
near the mass shell, i.e., in the nonperturbative region. At $x$ close to 1,
the situation is similar  [3]. As was shown in [1], eq.1 is true for the
real photon at $0.2< x < 0.7$. With $\mid p^2 \mid$ increasing the
applicability region of (1) expands.

In  expression (1) for the structure function  $F_2(x,p^2,Q^2)$  the
evolution of $F_2(x,p^2,Q^2)$  with $Q^2$ was not accounted, because for the
solution of the evolution equations it is necessary to know $F_2$ throughout
the whole interval of $x$. That it why one may expect that (1) is true at
intermadiate values $Q^2\sim 5 - 10 GeV^2$, where the role of evolution is
still inessential. (When comparing with the experiment it turned out that
(1) well describes experiment even for $Q^2 \approx 20 GeV^2$).

In this paper we will try to partially correct this disadvantage, to take
into account the evolution equations and to calculate gluonic distribution
in the real and virtual
photon. The calculation of gluonic distribution is preferable by virtue of
our approach because gluons are produced by quarks and  gluon energies
are essentially smaller than quark energies. This allows us to hope (and is
confirmed by calculation) that basing on the known  quark distributions
in the photon at $x > 0.2$ we shall be able to find gluonic distribution at
noticeably smaller $x$. (The $x > 0.8$ region weakly affects gluonic
distribution at $x < 0.5$). The interest to gluonic distribution in the
photon is also connected with the proposal to measure it on HERA
accelerator [8].

Calculations will be performed in the leading (LO) and next to  leading (NLO)
approximation. We consider two variants of the boundary conditions.

The first  variant is based on an assumption that all gluonic distribution in
the photon stems only from their emission by quarks and that the photon has
no intrinsic gluons of nonperturbative origin. I.e., we will believe that at
some low normalization point $Q^2=Q^2_0$ the gluonic distribution (as well
as the sea quark distribution -- we determine them in Sec.3) vanishes. This
idea is close to the forwarded earlier analogous idea on the origin of
gluons and sea quarks in hadrons (see, e.g. [9]).  It seems that in the
case of a photon, espesially of a virtual one, such an idea has much more
rights for existence than in the case of hadrons. Let us once more emphasize
that the main difference between this consideration and the previous ones
(see, e.g. [10--20]) is that the hadronic part of the structure function of
a photon at $Q^2 = Q^2_0$ is taken  from [1] and the point--like photon is
separated from its hadronic components in a way as it was explained above.

The second variant of the boundary conditions is also related to the results
of [1]. The second (hadronic) term in eq.(1) corresponds to the
contribution of vector meson, i.e. to transition of a photon into vector
meson with subsequent scattering of the projectile virtual photon on this
vector meson (see Fig.2). Therefore one may speak about  gluonic sea of
the photon which corresponds to the vector meson intrinsic gluons
(naturally, with a corresponding normalization).

As follows from our results, the gluonic distribution appears to be
essentially different in these two variants. Thus, comparison of gluonic
distribution with experiment would clarify hadronic structure of photon.

In Sec.2 we discuss the evolution equations in the NLO approximation. In
Sec.3 we will find their solution and discuss the results obtained.

\newpage
\vspace{5mm}
{\bf 2. ~\underline{Evolution Equations. Partial Solution}}

.~~~~{\bf \underline{of Inhomoqeneus Equation.}}

\vspace{3mm}
As has been originally shown by Witten [10], evolution of the photon
structure function  differs from the standard evolution of the structure
functions of hadrons. The reason for this is that in the photon case there
exists an additional mechanism of direct (point--like) production of quarks
from photon (as well as of gluons in the next order in $\alpha_s$). This
mechanism results in logarithmic increasing of the structure function with
$Q^2$ unlike evolution of hadronic structure functions where quarks and
gluons may be produced only from  quarks or gluons. Formally, in the OPE
technique the moment expansion of the photon structure function has the form
[10]

$$F^n_2 (Q^2)\equiv\int^{1}_{0}dx \cdot x^{n-2}~F_2(x,Q^2)=\sum_{i=NS,S,G}
\hat{C}^i_n(Q^2/\mu^2, g(\mu^2))<\gamma \mid \hat{O}_{n}^i \mid \gamma >+$$
\be
+C^{\gamma}_n(Q^2/\mu^2, g(\mu^2))< \gamma \mid O^{\gamma}_n \mid
\gamma >
\ee
Here $\hat{O}^i_n = (O^{NS}_n, O^S_n, O^G_n) \equiv \hat{O}_n$,
correspondingly, nonsinglet and singlet quark and gluon operators of spin
$n$ and twist-2 and $O^{\gamma}_n$ are the photon operators (see e.g.[11]),
$\hat{C}^i_n(Q^2/\mu^2, g^2(\mu)$) is the column of Wilson coefficients

$$
\hat{C}^i_n(Q^2/\mu^2, g(\mu))=
\left [ \begin{array}{l}
C^{NS}_n  (Q^2/\mu^2, g(\mu)) \\
C^{S}_n  (Q^2/\mu^2, g(\mu))\\
C^G_n  (Q^2/ \mu^2, g(\mu))
\end{array}
\right ] \equiv \hat{C}_n(Q^2/\mu^2,g(\mu))
$$
As was noted in [10], the second term in (2) for the case of hadrons would
be the $O(\alpha)$ correction, since $ C^{\gamma}_n(Q^2/\mu^2, g(\mu) \sim
O(\alpha)$. For the case of photons this term should be also taken into
account because matrix element $<\gamma \mid \hat{O}^i_n \mid \gamma >$ in
the first term of (2) is also of order $\alpha$ ($< \gamma \mid O^\gamma_n
\mid \gamma >$ is, naturally of order 1).

Thus, the renormalization group equation for $C_n$ takes the form [10]:

\be
(\mu \frac{\partial}{\partial\mu} + \beta(y) \frac{\partial}{\partial y})
\left [
\begin{array}{l}
\hat{C}_n\\
C^{\gamma}_n
\end{array}
\right ] =
\left (
\begin{array}{ll}
\hat{\gamma}_n & O\\
{\bf K}_n & O
\end{array}
\right )
\left [
\begin{array}{l}
\hat{C}_n\\
C^{\gamma}_n
\end{array}
\right ]
\ee
where $\hat{\gamma}_n$ is the standard matrix of anomalous dimensions and
${\bf K}_n = (K^{NS}_n, K^S_n, K^G_n)$ leads to mixing of photonic and
hadronic operators. One can find the explicit form of $\hat{\gamma}_n$ and
$K$ up to terms $\alpha_s ln Q^2$ in refs.[12,13]. (See also
expressions 8-10 ). The term ${\bf K}_n$ gives an additional contribution
in the Gribov-Lipatov-Altarelli-Parisi (GLAP) evolution equation to the
quark distribution function from the direct quark and gluon production by
photon (see Fig.3). \footnote{It should be reminded that  $\bf{K}$ play the
role of anomalous dimensions and are $p^2$ independent. All $p^2$ dependence
in (2) is gathered into matrix elements $< \gamma \mid \hat{O}^i_n \mid
\gamma >$ and appears in fact as a boundary condition.} Solving eq.(3) and
substituting the result into (2) we can find the final expression for
$F_2(Q^2)$. In symbolical notation it looks like (see, e.g. [10, 12]):

\be
F^n_2(Q^2) = \hat{C}_n(1,\bar{g}^2) \hat{M}^n (\bar{g}^2,
\bar{g}^2_0) < \gamma \mid \hat{O}_n (Q^2_0) \mid \gamma > +
\ee
$$
\hat{X}_n(\bar{g}^2, \bar{g}^2_0) \hat{C}_n(1, \bar{g}^2) +
C^{\gamma}_n(1, \bar{g}^2)
$$
where

$$
\bar{g}^2 = 4 \pi \alpha_s(Q^2) ; ~~ \bar{g}^2_0 = 4 \pi
\alpha_s(Q^2_0);
$$

$$\hat{M}^n (\bar{g}^2, \bar{g}^2_0) = exp \int\limits_{\bar{g}}
^{\bar{g}_0} dg \frac{\hat{\gamma}_n(g)}{\beta(g)}$$

$$
\hat{X} = \int \limits _{\bar{g}} ^{\bar{g}_0} dg \frac
{{\bf K}_n(g)}{\beta(g)} \cdot \hat{M}^n(\bar{g}^2, g^2)
$$

Functions
$$C^{S, NS}_n(1, \bar{g}^2) =(1 + \frac{\bar{g}^2}{16 \pi^2} B^{\psi}_n
{}~~~) \cdot
\left \{
\begin{array}{l}
1 ~~ \mbox{for}~~ NS\\
<e^2> ~ \mbox{for}~~ S
\end{array}
\right.$$
\be
C^G_n(1, \bar{g}^{2}) = <e^2> \frac{\bar{g}^2}{16 \pi^2} B^G_n
\ee
$$C^{\gamma}_n(1, \bar{g}^{2}) = \frac{\alpha}{4 \pi}~ 3 \sum_{i=1}^f e^4_i~
B^{\gamma}_n;~~~ <e^2> = (1/f) \sum_{i=1}^f e^2_i$$
$B^{\gamma}_n = (2/f)~B^G_n$ and $B^G_n$ are presented in [12].

In the general case $\hat{X}^n, \hat{M}^n, \hat{\gamma}_n, {\bf K}_n$ -- are
matrices. In what follows all calculations are made on an example of
nonsinglet part of $F_2$ : $F_2^{NS}$. In this case $M^n, \gamma_n,
K_n$ are algebraic expressions. (The calculation of the singlet
part is principally the same.)

For the nonsinglet case (4) is reduces to:

\be
F_2^{n(NS)}= C^{NS}_n (1,\bar{g}^2)~\{ M^n_{NS}(\bar{g}^2, \bar{g}^2_0)
 <\gamma \mid O_n^{NS}(Q^2_0) \mid \gamma > + X_n^{NS}(\bar{g}^2,
\bar{g}^2_0) \}
\ee
The factor in the curly brackets is usually treated as the moment of
distribution function (see, however [21],  when the other
difinition of the distribution functions was discussed, in which $C^{NS}_n$
also was included in it). In the leading logarithmic approximation $C^i_n=1$
and the standard definition $F^n_2= \sum e^2_i q^i_n,~~ q^i_n=\int
{}~x^{n-1}q_i(x)dx$ is reproduced. In NLO the moments of quark distribution
function $q^{NS}$ are defined by

$$q^{NS}_n= \int x^{n-1}~ q^{NS}(x) dx=M^n_{NS} (\bar{g}^2, g^2_0) <\gamma
\mid O^{NS}_n \mid \gamma > + X^{NS}_n(\bar{g}^2, g^2_0)$$
In the same way the moments of singlet and gluon distribution
functions are defined. The first term in (4)  corresponds to the
contribution from $C^{NS,S,G}_n(Q^2/\mu^2, g^2)$  in (2) (with account of
the evolution (3)), the second and the third terms arise from
$C^{\gamma}_n (Q^2/\mu^2, g)$ in (2) as a result of solution of eq.(3).
Usually, the first term in (4) is treated as hadronic part of the structure
function, since its form coincides with the solution of renormalization
group equation for hadrons; the second term -- as "point--like" production
of quarks and gluons by photon (which involves characteristic $ln~Q^2$
behaviour) and  the third term -- is the proper photonic part which is
included into quark distributions in some factorization schemes (e.g.,
$DIS_{\gamma}$, in ref.[16]). However, such separation may be dangereous
because of the possibility of double counting, as was discussed in [1]. One
can make this separation correctly basing on approach and results [1] where
it is performed in the framework of QCD sum rules. To this end it is more
convenient to use the evolution equation language.

The explicit form of the evolution equations for the distribution functions
$q_i(x)$  and $G(x)$ was given in [14] in the leading logarithmic
approximation while the next order terms were taken into account in [11,15].
These equations have the form: ($i$ means flavours of quarks)

\be
\frac{dq_i}{dt}=\frac{\alpha_s}{2\pi}(P_{qg} \oplus q^i + P_{qq} \oplus
G)+ \frac{\alpha}{2\pi}K^i_q \oplus \Gamma ,
\ee
$$\frac{dG}{dt}=\frac{\alpha_s}{2\pi}~(P_{gq}\oplus \sum_i(q^i+\bar{q}^i)+
P_{gg}\oplus G)+\frac{\alpha}{2\pi}~K^G \oplus \Gamma $$
where $t=ln~Q^2/\Lambda^2;~ P=P^0+\frac{\alpha_s(Q^2)}{4\pi}P^1$, and
$P$--are the splitting functions [11-13],

$$K^i_q = (K^{i,0}_q +  \frac{\alpha_s}{2\pi}K^{i,1}_q) ~~~~~~~
K^G(x)=\frac{\alpha_s}{2\pi}~K^1_G (x)~~~~~~~~~(7a)$$
Here $P \oplus q$ stands for convolution

$$\int \limits_{x}^{1} \frac{dy}{y} P(x/y) q(y,t),$$
$\Gamma(y) = \delta(1-y)$ is the photon distribution function in the
photon. $\alpha/2\pi K_{q,G}(x)$ are related with anomalous dimensions
${\bf K}_n = (K^{NS}_n, K^S_n, K^G_n)$ by
Mellin transformation.  The exact relation can be found in paper [11]. We do
not present it here since in what follows we shall need only $K_{q,G}(x)$,
the explicit form of which is [11] :

\be
K^{i(0,1)}_q(x) = 3 e^2_i ~  \kappa^{(0,1)}_q(x), ~~~i = u,d,s
\ee
where $\kappa^0_q(x) \equiv \kappa(x)$ is defined in (1);

\be
\kappa^1_q(x) = \frac{2}{3} e^2_i \{ 4-9x - (1-4x)ln x - (1 -2x)ln^2x +
4ln(1-x)+
\ee
$$+[4ln x - 4ln x ln(1-x) + 2ln^2x - 4ln(1-x) + 2ln^2(1-x) - (2/3) \pi^2+
10]\kappa(x)\}$$
$K^1_G(x)$ was taken from [16] where the error in [12] and in the papers
based on it was corrected

\be
K^1_G(x) = \sum_{i} e^2_i \cdot 4[-16 + 8x + \frac{20}{3} x^2 +
\frac{4}{3x} - (6 + 10x)ln x - 2(1+x)ln^2 x]
\ee
Here, analogously to (6), the function $F_2$ is determined through quark and
gluonic functions as [15, 11]

$$F_2(x) = x \left \{ \sum_{i} e^2_i(\hat{q}_i(x) + \frac{\alpha_s}{4 \pi}
B_{\psi} \oplus \hat{q}_i)+ \right.$$ \be \left.+ <e^2> \cdot
\frac{\alpha_s}{4\pi} B_G \oplus G + 3 \sum_{i} e^4_i
B_{\gamma}(x)\alpha/4\pi \right \}
\ee
where $\hat{q} = q + \bar{q}$

$$B_{\psi}(x) = (8/3)~\left \{\frac{9+5x}{4} - \frac{1+x^2}{1-x} lnx -
\frac{3}{4} \left [~\frac{1+x^2}{1-x} \right ]_+ +\right.$$
$$\left.+ (1+x^2) \left [~\frac{ln(1-x)}{1-x} \right ]_+
- (9/2 + \pi^2/3) \delta (1-x)~\right \}$$
$$B_G(x) = 2f \left ( (1-2x+2x^2)~ ln \frac{1-x}{x} - 1 +
8x(1-x) \right );$$
$$B_{\gamma}(x) = (2/f) \cdot B^G(x))$$
$B^{\psi}_{n}$ and $B^{\gamma}_n$ defined in (5) are $n$- moments of these
functions. (All expressions here are in the $\overline{MS}$
scheme). The last terms in eqs.(7) correspond to the process of point-like
(direct) production of quarks and gluons from photon in LO and NLO: we
preserve only the terms proportional to $ln Q^2$ or $\alpha_s ln Q^2$ (see
Fig.3).

In the case of $DIS_\gamma$ scheme suggested in Ref.[16]

$$F^{DIS_{\gamma}}_2 (x) = x \left \{\sum_{i}~ e^2_i \left (\hat{q}^
{DIS_{\gamma}}(x) + \frac{\alpha_s}{4 \pi}~ B_{\psi} \oplus~
\hat{q}^{DIS_{\gamma}}(x) \right ) + \right.$$
$$\left.+ < e^2 > \frac{\alpha_s}{4
\pi} B_G \oplus G(x) \right \}~~~~~~~~~~~~~~~~~~~~~~~~~~~~~(11a)$$
where the last term in (11) is included in the
$\hat{q}^{DIS_{\gamma}}(x)$. This redefinition leads to transformation of
$K^{(1)}_{q,G}( \equiv K^{i,1}_q$  or $K^1_G$, see 8-10).

$$K^{(1)~DIS_{\gamma}}_{q,G} = K^{(1)}_{q,G} ~ + ~ \delta K_{q,G}$$
where $\delta K_{q,G}$ from eq.(3.1, A.11--A.12) in ref.[16]. This scheme
has many advantages in comparison with the $\overline{MS}$ scheme (see
discussion in [16]).

Especially in the $DIS_{\gamma}$ scheme there is no strong difference
between LO and NLO results. Hereafter we will work with the $DIS_{\gamma}$
scheme (and omit $DIS_{\gamma}$ indeces). We would like to stress, however,
that our results are comparatively insensitive to the choice of
renormalization scheme ($\overline{MS}$ or $DIS_{\gamma}$), see discussion
at the end of the paper.

The evolution equations (7) are equivalent to the renormalization group
equations for the moments (3). The form of equations (3) and (7) is
independent of the photon virtuality, the $p^2$ dependence is hidden in the
matrix elements $<\gamma \mid \hat{O}_n \mid \gamma >$ and therefore appears
only through the boundary conditions. Evidently, for virtual photon the
renormalization group equations are valid at $Q^2 >> -p^2$. In what follows
we consider only the case of transversally polarized real or virtual photon
and for simplicity the notation $F_2$ will be used for $F^T_2$ at $p^2
\not=0$. The unpolarized virtual photon case will be discussed at the end of
Section 3.

The evolution equations (7) are inhomogeneous. As is known, the general
solution of inhomogeneous equation is given by the sum of partial solution
of inhomogeneous equation plus general solution of homogeneous equation. The
latter coincides with the standard solution of evolution equation for
hadrons. We choose the partial solution of evolution equation in NLO by
requiring its vanishing at $Q^2 = Q^2_0$ (see refs.[15-17]). The solution of
inhomogeneous equations, represented in terms of moments of singlet and
gluon distributions is given by [16]

$$ \left [
\begin{array}{l}
q^s_n\\
G_n
\end{array} \right ]
= \frac{\alpha}{2 \pi} ~ \frac{2}{\beta_0} \left \{ \left [2\pi/\alpha_s~
\hat{P}_+  -\frac{2}{\beta_0} \hat{P}_+ \hat{P}_1
\hat{P}_+ + \right.\right.$$
$$\left.+ \frac{\beta_1}{\beta^2_0} \lambda_+ \hat{P}_+ +
\frac{\hat{P}_- \hat{P}_1 \hat{P}_+}{\lambda_+ - \lambda_- - \beta_0/2}~\right
]
\times $$
$$\times \frac{1 - L^{1-2 \lambda_+/\beta_0}}{1-2
\lambda_+/\beta_0} ~ \hat K_0 - \frac{1 - L^{-2 \lambda_+/\beta_0}}{2
\lambda_+/\beta_0} \cdot \left [ \hat{P}_+~\hat{K}_1 - \frac
{\beta_1}{2 \beta_0} \hat{P}_+ \hat{K}_0 - \right.$$
\be
\left.+ \frac{2}{\beta_0}
\left ( \hat{P}_+ \hat{P}_1 \hat{P}_+ - \frac{\beta_1}{2\beta_0}~ \lambda_+
\hat{P}_+ - \frac{\beta_0}{2} ~ \frac{\hat{P}_+ \hat{P}_1
\hat{P}_-}{\lambda_- -  \lambda_+ - \beta_0/2}\right ) \hat{K}_0 \right ] +
\ee
$$\left.+
(\mbox{terms with interchange}~~~ \lambda_+ \rightarrow \lambda_-, \hat{P}_+
\rightarrow \hat{P}_-) \right \} $$
Here $\hat{P}_1, \hat{P}_+, \hat{P}_-$
are matrices; $q^s_n = \sum_{i=1}^{f}~ q^i_n + \bar{q}^i_n$

$$\hat{P}_1 =
\left ( \begin{array}{ll}
P^1_{qq}, & P^1_{qg}\\
P^1_{gq} & P^1_{gg}
\end{array}
\right ), ~~~
\hat{P}_{\pm} = \pm \frac{\hat{P}_0 - \lambda_{\mp}
\cdot I}{\lambda_+ - \lambda_-}, ~~~ \hat{P}_0 =
\left ( \begin{array}{ll}
P^0_{qq} & \tilde{P}^0_{qg} \\
P^0_{gq} & P^0_{gg}
\end{array}
\right )$$

$$\lambda_{\pm} = 1/2(P^0_{qq} + P^0_{gg} \pm \sqrt{(P^0_{qq}-P^0_{gg})^2 +
4 \tilde{P}^0_{qg}~P^0_{gq})}, ~~~ \tilde{P}^0_{qg} = 2f P^0_{qg},$$

$$\hat{K}_0 = {K^0_q \choose 0}, ~~~ \hat{K}_1 = {K^1_q\choose K^1_G}, L =
\frac{\alpha_s(Q^2)}{\alpha_s(Q^2_0)};~~~K^{(0,1)}_q = \sum_{1=1}^{2f}
K^{i(0,1)}_q$$
$P^{0,1}_{qq},~ P^{0,1}_{qg}$ etc. mean the n-th moment of
the standard splitting functions.

As mentioned in refs.[15,16], such a choice of the partial solution when
even at large $Q^2$ the terms $L^{-2\lambda_\pm /\beta}$ are retained
permits one to get rid of
nonphysical poles at  $n = 1.596$ and $n=2$ corresponding to $1 -
2\lambda_+/\beta_0 = 0$ and $\lambda_+ = 0$. These poles would correspond to
the poles in x-distribution of the type $1/x^{1.596}$ and $1/x^2$. The
vanishing  at the same points numerators result in the logarithmic
divergences in x-distributions instead of power-like ones. The partial
solution chosen in this way is completely analogous to the term
$\hat{X}(\bar{g}^2, g^2_0)$ in the solution of eq.3. As already has been
mentioned, this term incorporates the terms proportional to $ln Q^2$,
related to the point-like mechanism of quark and gluon production by photon
in LO and NLO. However, it would be erroneous to think that the solution of
inhomogeneous equation with zero boundary conditions takes into account the
\underline{whole} contribution from the process of the point-like quark
production at a given $Q^2$ while the general solution of homogeneous
equation (equivalent of $M^n$ in (4)) - only~ hadronic~ part,~ i.e. the~
contribution~ of~ soft~ and~ collinear \\quarks \footnote{Such a
distinction may be achieved, as we shall see in what follows, but only at
certain boundary conditions.}.  At such a choice, there may appear a double
counting (or a deficit if we take hadronic part to be zero [16]).

One of the methods to solve this problem correctly was discussed in the
Introduction. This is to use as boundary conditions the results of ref. [1]
- eq.(1),, where the point-like and hadronic contributions into photon
structure functions -- respectively, the first and the second terms in curly
brackets in (1) -- are calculated  in QCD. In such case the mechanism of the
point-like production corresponds to solution of inhomogeneous equation with
the boundary conditions from the first term of (1) to which one should add
the hadronic part corresponding to solution of homogeneous equation with the
boundary conditions from the second term of (1).

Owing to linearity of the evolution equations, it is, naturally, more
convenient to take the solution of inhomogeneous equation with the zero
boundary conditions  and to add the solution of homogeneous equation with
the boundary conditions which are the sum of the boundary conditions of the
point-like and hadronic parts. But in doing so, the solution of homogeneous
equation is not \underline{only} the contribution of hadronic part (e.g., it
is, erroneous to choose in this case the boundary condition
from VDM considerations).

Before going to solution of homogeneous equation, let us come back once more
to the general form of inhomogeneous equation (7).

Consider two functions $\varphi(x) \equiv q^{NS}(x) = (1/2)(u(x) - d(x))$ and
$\tau = -(1/2)(u/e^2_u - d/e^2_d)$ where $e_i$ is the charge square of the
corresponding quark. The evolution equation (7) and the boundary condition
on $F^T_2$ are the same for $d$ and $s$ quarks besides 
small difference in the boundary condition  (1) arising from the mass
difference of $\rho$ and  $\varphi$ mesons. (The calculation shows that the
influence of this difference on gluon distribution is very small)

Thus, we may safely take $d(x)=s(x)$, and, correspondingly,

\be
d(x)=s(x),~~~~~~~~\hat{q}_i \equiv (q_i+\bar{q}_i)=4\varphi \frac{ e^2_i}
{\Delta} + \frac{4 e^2_u e^2_d}{\Delta}
\ee
where $\Delta=e^2_u -e^2_d=1/3$. The inhomogeneous evolution equation (7)
expressed through these functions is:

$$\frac{d\varphi
(x,t)}{dt}=\frac{\alpha}{2\pi} \frac{1}{2}(e^2_u-e^2_d)3 \left ( k^0_q(x)+
\frac{\alpha_s(t)}{2\pi} k^{(1)}_q(x) \right ) + $$
$$+ \frac{\alpha_s(t)}{2\pi}~ \int\limits^1_x
\frac{dy}{y}P_{qq}(x/y)\varphi(y,t)$$
\be
\frac{d\tau(x,t)}{dt}=\frac{\alpha_s(t)}{2\pi}~\int\limits^1_x \left (
P_{qq}(x/y)\tau(y,t)+P_{qg} (x/y)\frac{1}{2}~\frac{\Delta}{e^2_u
e^2_d}~G(y,t) \right ) \frac{dy}{y} \ee $$\frac{dG(x,t)}{dt}=\frac{\alpha
\alpha_s(t)}{(2\pi)^2}~K^1_G(x)+\frac{\alpha_s(t)}{2\pi}~\int\limits^1_x~
\left [ P_{q G}(x/y)(\frac{4\Sigma
e_i^2}{\Delta}\varphi(y,t)+\right.$$
$$\left.+ \frac{4 f e^2_u
e^2_d}{\Delta} \tau(y,t)) +P_{gg}(x/y)G(y,t) \right ]~\frac{dy}{y}$$
where notations are the same as in (7), (7a), (8). The
evolution equation for  $\tau(x)$  has the form of standard evolutions for
hadrons and all quark and gluon point--like production is now absorbed in
$\varphi$ function. Thus, $\varphi(x)$  may be treated as  valence quark
distribution in the photon and $4(e^2_u e^2_d/\Delta)\tau(x)$ as "sea
quark" distribution \footnote{Such detemination of "valence" and "sea"
quarks distribution is close to that used in [20].}.  In the next Section we
consider two variants of boundary conditions and discuss the obtained
solutions.

\vspace{5mm}
3.~~\underline{Solution of Evolution Equations. Discussion of Results.}

\vspace{3mm}
In the first variant of the boundary conditions we assume that at some,
relatively small $Q^2=Q^2_0$  there are no $\tau$  ("sea") quarks, as well
as gluons but there are only "valence"  quarks ($\varphi$--quarks).

Physically, this corresponds to assumption that both gluons and sea quarks
in photon are produced in perturbative way due to their emission by
valence quark (as well by a bare photon for the gluon case) and they are
absent at the low normalization point $Q^2=Q^2_0 \sim 1 GeV^2$. As has been
already noticed in the Introduction, an analogous idea was forwarded to
describe the structure function of hadrons where it was supposed [9], that
at small $Q^2=Q^2_0 \sim 1 GeV^2$ a hadron consists only of valence quarks.
But such an idea failed in description of the hadron structure functions --
it appears  to be inconsistent with experiment. This idea has better
chances for photon since it can be naturally thought that photon has no
gluons and sea quarks (in the  sense defined at the end of the previous
Section) as their constituents. Note that for virtual photon $Q^2$ should be
chosen larger than  $\mid p^2 \mid$. As the hadronic component contribution
into structure function (1) decreases rapidly with  $\mid p^2 \mid$
increase, and hence it follows that for strongly virtual photon our
requirement is fulfilled automatically.

The total solution for the quark distribution function consists of the sum of
solution of inhomogeneous equation which was under discussion in the
previous section and of the solution of homogeneous equation

$$\frac{d\varphi(x,t)}{dt} =\frac{\alpha_s(t)}{2\pi}~\int\limits^1_x
{}~(P_{qq}(x/y)~\varphi(x,t))~\frac{dy}{y}$$
$$\frac{dG(x,t)}{dt}=\frac{\alpha_s(t)}{2\pi}\int\limits^1_x \left \{
[P_{qG}(x/y)(\frac{4\Sigma e^2_i}{\Delta}\varphi(y,t)
+\right.$$
$$\left.+ \frac{4f e_u^2
e^2_d}{\Delta}\tau(y,t))]+P_{gg}(x/y)G(y,t) \right \}$$
\be
\frac{d\tau(x,t)}{dt}=\frac{\alpha_s(t)}{2\pi}
\int\limits^1_x~(P_{qq}(x/y)~\tau(y,t)
+\frac{1}{2}P_{qq}(x/y)\frac{\Delta}{e^2_u e^2_d}~G(y,t))\frac{dy}{y}
\ee
with the boundary conditions from (1). Since we restrict ourselves to NLO of
inhomogeneous equation i.e., to terms $\alpha_s ln Q^2$, then in the same
order the homogeneous equation can be solved in the leading
logarithmic approximation, so $\frac{\alpha_s(t)}{2\pi}=\frac{2}{\beta_0t}$.
In this approximation in eq.(11a) for $F_2(x,Q^2_0)$ (note, that
$G(x,Q^2_0)=0$)

\be
(1/x)~F_2(x,Q^2_0)=\sum~e^2_i~[~\hat{q}_i(x)+\frac{\alpha_s}{4\pi}~B_{\psi}~\oplus
{}~\hat{q}_i(x)~]
\ee
one should omit the term ($\alpha_s/4\pi) ~B_{\psi}~\oplus~\hat{q}_i$ ,
since the solution of homogeneous equation $\hat{q}_i(x,Q^2)$ has no $ln
Q^2$ and the contribution from the solution of inhomogeneus equation at
$Q^2=Q^2_0$ is zero. As the estimates show, this term gives a small
contribution to the gluonic structure function at $x < 0.5$.  Therefore the
boundary conditions have the form

\be
F_2(x,Q^2_0)=x \sum e^2_i q_i (x,Q^2_0)
\ee
$$G(x,Q^2_0)=\tau(x, Q^2_0)=0 $$
Accounting for (1), (13) we have

\be
\varphi(x,t_0)=\frac{\alpha}{4\pi}~\left \{ ~[2+\kappa(x)(ln~\frac{Q^2_0}
{x^2(s_0-p^2)}-3+\frac{p^2}{s_0-p^2})]+ \right.
\ee
$$\left.+ \frac{s^2_0}{2(m^2_{V}-p^2)^2}~\rho(x) \right \}$$
where  $t_0 = ln(Q^2_0/\Lambda^2)$
$$\rho(x)=\kappa(x)-\frac{8\pi^2}{27 x^2 s^2_0}<0 \mid \frac{\alpha_s}{\pi}
G^2_{\mu \nu} \mid 0 >~~~~~~~~~~~~~~~~~~~~~~ (18')$$
and $m_V$ means the weighted average of $\rho$ and $\varphi$ mass. Let
us represent $\varphi(x, t)$ as a series

\be
\varphi(x, t) = \varphi(x, t_0) + \sum _{1}^{\infty} \frac{\varphi_m
(x)}{m!}(ln \frac{t}{t_0})^m
\ee

For $\tau(x, t)$ and $G(x, t)$ similar
expressions take place:

\be
\tau(x,t) = \sum_{1}^{\infty} \frac{\tau_m(x)}{m!}(ln\frac{t}{t_0})^m,
{}~~~ G(x,t) = \sum_{1}^{\infty} \frac{G_m(x)}{m!}(ln \frac{t}{t_0})^m
\ee
Substituting (19),(20) into the evolution equations we get recurrent
relations $(f = 3)$

$$
\varphi_m(x)=(\frac{2}{\beta_0})^m
\underbrace{P^0_{qq} \oplus P^0_{qq} \oplus....P^0_{qq}}_{\mbox{$m$~
times}} \oplus \varphi(x, t_0) $$
$$ G_{m+1}(x)=\frac{2}{\beta_0} \left [ \frac{4}
{\Delta}~\sum e^2_i P^0_{gq}~\varphi_m(x) + \frac{12 e^2_u
e^2_d}{\Delta}~\tau_m (x) \oplus P^0_{gq} + P^{0}_{gg} \oplus G_m (x)\right
]$$ $$\tau_1(x) = 0$$
\be
\tau_{m+1}(x) = \frac{2}{\beta_0} [P^0_{qq} \oplus
\tau_m(x) + \frac{1}{2} \frac {\Delta}{e^2_u~e^2_d} P_{qg} \oplus G_m(x)]
\ee
In
order to solve equations (21) it is necessary to know the function
$\varphi(x, Q^2_0) \equiv \varphi(x,t_0)$ in the region $0.75 < x < 1$ where
eqs.(1) and (18) are invalid. Notice that the uncertainty at large $x$
arises only in the second term in (1) or (18) related to the
vector meson contribution.  The natural extrapolation comes from quark
counting rules and corresponds to $f_{\rho}(x) \sim (1-x)$ at large $x$. We
considered also the other forms of extrapolation $f_{\rho}$ in the domain of
large $x$:  $f_{\rho} (x) = A(1-x)^{\alpha}$, where $A$ was determined from
the matching with the second term in (1) at the point $x = 0.75$. It was
found that $G(x,Q^2)$ at $x \leq 0.5$ varies only by a few per cent if
$\alpha$ varied from $1$ to 2.
We emphasize, that the account in the evolution equations of singular terms,
proportional to $\delta(1-x),~\delta'(1-x)$  etc, (see [1]) would be a
mistake. These terms (e.g. the term, proportional to operator
$\alpha_s~\bar{\psi}~\Gamma_i~\psi~\bar{\psi}~\Gamma_i~\psi$)  formally
appear in the OPE over the soft photon virtuality $1/p^2$. However, as was
explained in ref.3, these terms are unphysical because they correspond to
the bump at $x=1$ and must be compensated in the correct theory.

The series (19),(20) rapidly converge at $x \leq 0.6$ and $Q^2 < 100 GeV^2$,
so it is sufficient to take into account three terms in the expansion. In
the homogeneous equation we can put the one-loop expression for $\alpha_s$:
$\alpha_s = 4 \pi/\beta_0 ln(Q^2/\Lambda^2)$, $\Lambda = 230 MeV$, since this
equation is treated in the leading logarithmic approximation. In the case of
inhomogeneous equation (10) the calculations are performed in NLO in
$DIS_{\gamma}$ regularization scheme and, correspondingly, two-loop
expression for $\alpha_s$ is used:

$$\frac{1}{4\pi} \alpha_s(Q^2) = \frac{1}{\beta_0 ln(Q^2/\Lambda^2)} -
\frac{\beta_1}{\beta^3_0}~ \frac {ln~
ln(Q^2/\Lambda^2)}{ln^2(Q^2/\Lambda^2)}$$
The functions $q(x),~ G(x)~~ (q(x) \equiv q^s(x)$ is the
singlet quark distribution) were obtained from the solution of the
inhomogeneous equation (12) by the inverse Mellin transformation.
The results are presented in Fig.4--8. Dotted lines correspond to gluon
condensate contribution, thin solid lines -- to hadronic part and thick
lines -- total gluon distribution $x~G(x)/\alpha$. For $p^2\not=0$
(Fig.$5\div 7$)  thick solid lines correspond to transversally polarized
photon case, and dashed lines -- unpolarized photon case, which will be
discussed at the end of the paper.

As is
seen from Figs.4-7 the negative gluonic condensate contribution is essential
only at $x < 0.1$, it increases steeply with $x$ decreasing and for the real
photon at $Q^2 = 5 GeV^2$ comprises a half of the total $G(x)$ at $x \approx
0.02$.  This means that the QCD sum rule approach prediction for the gluon
distribution function in the real photon is reliable up to $x > 0.02-0.05$
(in the framework of the chosen boundary condition - variant I(HI)).The
gluonic condensate contribution decreases rapidly with increasing of photon
virtuality $-p^2$.

As was discussed in the Introduction, such an essential extension of the
applicability region towards small $x$ is possible for the gluon
distribution function only. For quark distribution the applicability region
starts at larger $x > 0.1$.

As is seen from Figs.4-7, the hadronic part  is rather large
for the real photon ($\sim 50\%$ at $Q^2 = 5 GeV^2$ and $30\%$ at $Q^2 = 50
GeV^2$) and decreases rapidly with $\mid p^2 \mid$ increasing.

An approach similar to ours for the case of the virtual photon was
developed  in refs.[15,17]. Here, however, the hadronic part was not singled
out from the boundary conditions and considered separately, as we did.

The dashed lines in Fig.4 at $Q^2 = 5$ and $50 GeV^2$ represent the results
of ref.[16], where the NLO evolution equations were solved with the zero
boundary conditions for inhomogeneous as well as for homogeneous equations.
As is expected, they are much more less than our results although relative
difference decreases with $x$ increasing. The difference is large even at
$Q^2 = 50 GeV^2$ what means that the domain of $Q^2 \sim 50 GeV^2$ is far
from asymptotics, where dominant are the terms in the solution of
inhomogeneous equations ($\sim lnQ^2$ and $\alpha_s ln Q^2$) independent of
the boundary conditions.

Our results for the gluon distribution function are weakly dependent on the
choice of the normalization point $Q^2_0$. For the real photon this can be
seen from Fig.8. With $Q^2$ or $\mid p^2 \mid$ increasing the $Q^2$
dependence becomes even weaker. The LO results are not too much different
from NLO: the difference is less than 10-15\% almost everywhere except for
the highest considered virtuality $\mid p^2 \mid = 2 GeV^2$ where it is
about 30\%-40\%. For this reason the LO results are not shown in the
Figures.

We turn now to the variant II(H2) of the boundary conditions discussed in the
Introduction. The second term in the photon structure function, eq.1,
corresponds to the target photon into vector meson transition with
subsequent projectile virtual photon-vector meson scattering. At such an
interpretation it is natural to assume that at low $Q^2$ there are
nonperturbative gluons in the photon and their distribution is determined by
gluon distributions in vector mesons. As was shown in ref.[1], the second
term in (1) is related to the transverse structure function $f^T_V(x)$ of
the fictitious vector meson $V$ built from one flavour quark $q$ with unit
charge

\be
\frac{4\pi}{g^2_V} f^T_V(x)/x = \frac{3}{2\pi}~~ \frac{s^2_0}{m^4_V} \rho(x)
\ee
where $\rho(x)$ is given by (18) and $g^2_V$ is the $\gamma-V$ transition
 coupling constant. (Experimentally, $g^2_V/4\pi = 1.27$.)~ $f^T_V(x)/x$ is
 equal to quark distribution,

 $$f^T_V(x) = x [q(x) + \bar{q}(x)]$$
 In a similar way we can define the gluon distribution $G^T_V(x)$ in
 $V$-meson. Assuming the standard form of $x$-dependence, $xG^T_V(x) \sim
 (1-x)^3$, we have

\be
 x~ G^T_V(x) = c~ \frac{3}{8\pi^2} ~ \frac{s^2_0}{m^4_V} ~g^2_V(1-x)^3
\ee
where the normalization constant $c$ is determined from the requirement that
gluons in $V$--meson carry 40\% of $V$-meson momentum, $c = 0.42$. The gluon
distribution in the virtual photon is related to $x G^T_V(x)$ by

$$x~ G^T_\gamma(x, Q^2_0) = \frac{4 \pi \alpha}{g^2_V} \Sigma~ e^2_q~
\frac{m^4_V}{(m^2_V-p^2)} x~G^T_V(x) =$$
\be
= c~ \frac{3\alpha}{2\pi}~\Sigma
e^2_q~ \frac{s^2_0}{(m^2_V-p^2)^2} (1-x)^3
\ee
Eq.24 is used as the boundary condition for the gluon distribution in the
variant II of boundary conditions. In the same way the boundary conditions
for sea quark distributions can be also imposed. But since they carry only
10\% of the vector meson momentum, their contribution to the final gluon
distribution is negligibly small. We emphasize that (24) is the boundary
condition for the hadronic part of the gluon distribution function, for
nonhadronic, point-like part the boundary condition is still zero. As
before, the boundary conditions  (17) are used for valence quark
distributions. (The account of the order $\alpha_s$ gluonic contribution to
$F_2(x,Q^2_0)$ in (17) is beyond the accuracy of our calculation and, as the
estimate shows, this contribution is small numerically).

The results of the calculations in the variant II are presented in
Figs.9-12. All notations are the same, as in Fig.$4 \div 8$.
In this variant the hadronic part dominates in the gluonic
contribution function for the real or virtual photon even at $Q^2 = 50
GeV^2$. In comparison with the variant I $G(x,Q^2)$ is essentially larger,
especially in the case of the real photon - by a factor of 2-3. The relative
difference of $G(x,Q^2)$ in variants I and II decreases with $Q^2$
increased as a result of increasing role of point-like gluon production. The
difference in gluon distributions in variants I and II is going down steeply
with increasing of the photon virtuality $\mid p^2 \mid$.

The results of ref.[18] at $Q^2 = 10 GeV^2$ are shown in Fig.13 by dotted
line. The method
of the calculations of ref.[18] is different from ours in three aspects:
i)low normalization point in inhomogeneous differential equation with the
zero boundary condition at $Q^2 = 0.3 GeV^2$ in [18] comparing with our
$Q^2_0 = 1 GeV^2$; ii) our boundary gluonic distribution (24) is much more
concentrated at small $x$ than that
used in ref.[18] -- $xG_{\gamma}(x) \sim x^{0.5}(1-x)^{\alpha}$, where
$\alpha=0.1$ at normalization point $Q^2_0=0.3 GeV^2$ and $\alpha \sim 1$
at $Q^2 \sim 5 \div 10 GeV^2$; iii) the difference in the boundary
conditions for valence quarks - eq.(17).

The change
of the normalization point to $Q^2_0 = 2GeV^2$ (instead of $Q^2_0 = 1
GeV^2$) decrease gluonic distribution in our calculation by
10-12\% (Fig.8b).  Therefore, the $Q^2_0$ dependence of the final gluon
distribution is weaker in the variant II than in the variant I.

The results of ref.[22] are shown by dashed line in Fig.13, and results of
[23] --by crosses.

The negative gluon condensate contribution to the boundary condition for
valence quarks (eq.17) results in flattering of $G(x)$ at small $x \sim
0.02-0.05$.
Result of fit, made in [20], are higher than ours (in all six choises of
their input) at low $x,$ because  their boundary condition for $G(x)$ is
$2.5 \div 5$ times larger, than ours. In Fig.14 we compare our results (II
variant, H2)  whith three sets of boundary conditions, offered by [20].
(Three other sets are even more larger and we don't show them).

In the considered above two variants of the boundary conditions (17,18) and
(18,24) the gluonic distributions in the real or weakly virtual photon
$(\mid p^2 \mid \leq 0.5 GeV^2$) are essentially different. It is impossible
to remove this difference by varying the normalization point $Q^2_0$ in the
first  variant of the boundary conditions. The experimental investigation of
the gluon distribution in photon can shed light on the problem if there
are or not intrinsic gluons in the photon.

The calculation of the whole $F_2$ structure function can be done in a
similar way. Like the gluon distribution, quark distribution functions are
given by the sum of partial solutions of inhomogeneous equations with
boundary conditions $q^{inh}(x,Q^2_0)=0$ plus general solution of
homogeneous equation $q^{hom}(x,Q^2)$  with boundary conditions, following
from (1). As  we are interested only in the terms, proportional to $ln~Q$
or $\alpha_s ln~Q$, so the gluon contribution and $\alpha_s$--corrections to
quarks contributions in (11a)  should be calculated in $ln~Q^2$ order. This
means, that this contributions may be accounted only in the solution of
inhomogeneous equations since the terms, proportional to $ln~Q^2$, arises
only there. So, if we write $F_2(x)$ in the form

$$F_2(x,Q^2) = F_2^{inh} (x,Q^2) + F_2^{hom}(x,Q^2),$$
where

$$F^{inh}_2(x,Q^2) = x \left \{~\sum_{i} e^2_i(\hat{q}_i^{inh} (x,
Q^2) + \frac{\alpha_s}{4 \pi}~ B_{\psi} \oplus \hat{q}^{inh}_i (x,Q^2) +
\right.$$
$$\left. + <e^2> \frac{\alpha_s}{4 \pi}~ B_G \oplus G(x, Q^2) \right \}$$
then

$$F^{hom}_2(x, Q^2_0) = x~ \sum_{i=1}^{f}~ e^2_i~ \hat{q}^{hom}_i (x,
Q^2_0)$$
and the boundary conditions for $\hat{q}^{hom} (x, Q^2_0)$ are the same as
(17) for the I variant(the II variant gives almost the same results and we
will not discuss it).

The results for $F_2$ are shown in Figs.15--18. (The notations are the same
as for the gluon distribution function, Figs.4-7). The results for the real
photon case (thick solid line in Figs.15,16) are compared with the
experimental data of $TPC/ 2\gamma$ [24,25], PLUTO [26-28] and OPAL [29]
collaborations. For comparison the results of [18] (dashed line) and [20]
(crossed line) are shown. One can see, that our results are in a good
agreement with the data at $0.2 < x < 0.7$ and $Q^2 = 5 \div 41 GeV^2$
where our approach is valid. At $x < 0.2$ the gluon condensate contribution
becomes too large and our results are unreliable. We should notice that, of
course, at $Q^2 \geq 40 \div 50 GeV^2$ the role of $c$-quarks is not
negligible at $x < 0.3$ and one has to take them into consideration.

The results of the structure function $F_2$ calculations for virtual
photon are presented in Figs.17,18. The thick solid line corresponds to the
transversally polarized photon case, the dashed one - to the unpolarized
virtual photon case (the latter will be discussed below).

Note that our results comparatively weekly depend on the renormalization
scheme ($M \overline{S}$ or $DIS_\gamma$), as it can be seen from Fig.19,
where our results are plotted in these two schemes. An approximate scheme
independence arises because the scheme difference in the solution of
inhomogeneous equation (see [16]) is compensted by the opposite difference
of the solution of homogeneous equation. The main reason for this
circumstance is that we have no separate boundry conditions for homogeneous
equation (which is independent of the solution of inhomogeneous equation and
is determined from any physical ideas like VDM), but have boundary condition
for  \underline{total} $F_2(x,Q^2_0)$.

Finally, let us discuss the nonpolarized virtual photon case. It is reduced
to trivial redefinition of boundary conditions for homogeneous equation
(17,18). Here we now should use $F_2 = F^T_2 + (1/2)~F^L, ~~
F^T_2$ being defined in (1) and $F^L_2$ can be found in [1]

$$F^L_2 = \frac{3 \alpha}{\pi}~ \sum e^4_i~ 4 x^2(1-x)~
\frac{p^2}{p^2 - \frac{m^2_q}{x(1-x)}}$$
$m_q$ -- is the quark mass.

Note that there is no contribution from gluon condensate in $F^L_2$
(see [1]).

The results for the unpolarized virtual photon case are about 20\% higher
than for transversally polarized one and are shown by dashed line in
Figs.5-7, 10-12 for gluon distribution and in Figs.17,18 for $F_2$
structure functions. For comparison, the results of [19] (disregarding any
heavy quark contribution) are shown in Figs.10,11,18 by crossed line.

One of the authors (A.G.O.) is thankful to the members of the ITEP
Theoretical Physics Department for their hospitality. This investigation was
supported in part by the International Science Foundation Grants M9H000 and
RYE000 and by the International Association for the Promotion of Cooperation
with Scientists from Independent States of the Former Soviet Union Grant
INTAS-93-283.

\newpage
\centerline{\underline{\bf Figure Captions}}

\vspace{10mm}
\begin{tabular}{lp{12cm}}
{\bf Fig.~1.} & Bare Diagram.\\
  & \\
{\bf Fig.~2.} & Hadronic contribution into structure function.\\
  & \\
{\bf Fig.~3.} & Some diagrams corresponding to quark and gluon direct
production.\\
  & \\
{\bf Fig.~4.} & Gluon distribution in the real photon (thick solid line);
thin solid line and dotted line -- respectively, the contributions
from hadronic part and from gluonic condensate, the latter taken with
opposite sign.  The results are given for $Q^2=5,10,50 GeV^2$. The boundary
conditions eq.s (17,18) -- I variant, correspond to the normalization point
$Q^2_0=1 GeV^2$.  The dashed line stands for the result of ref.[16].\\
  & \\
{\bf Fig.~5.} & Gluon distribution for virtually polarized photon ($p^2 =
-0.5 GeV^2, Q^2_0 = 1~ GeV^2$); thick and dashed lines, respectively -
transversally polarized and unpolarized virtual photon case, hadronic part
and module of gluon condensate contribution are shown by dashed and dotted
lines, respectively.\\
  &\\
{\bf Fig.~6.} & The same as Fig.5 for $p^2=-1 GeV^2;~ Q^2_0= 4 GeV^2; ~Q^2=50
GeV^2$.\\
  & \\
{\bf Fig.~7.} & The same as Fig.5 for $p^2=-2
GeV^2_0;~Q^2_0= 6 GeV^2;~Q^2=100 GeV^2.$\\
  & \\
{\bf Fig.~8.} & Gluon
distribution  function in the real photon at $Q^2=10 GeV^2$ and different
$Q^2_0~, Q^2_0=1 GeV^2$ -- thick solid line,
$Q^2_0=2$ -- thin solid line). a) Boundary conditions eqs. (17,18) -- I
variant. b) Boundary conditions eqs.(18,24) -- II variant.
\end{tabular}

\newpage
\begin{tabular}{lp{12cm}}
{\bf Fig.~9.} & Gluon distribution in the real photon for the case of the
boundary condition (24) and (18) for quarks - II variant of the boundary
conditions, $Q^2_0 = 1 GeV^2$. The notations are the same as
in Fig.4.\\
  &\\
{\bf Fig.~10} & The same as Fig.5 for the II variant (eqs.18,24). The
results of [19] are shown by the crossed line for comparison.\\
  &\\
{\bf Fig.~11} & The same as Fig.6 for the II variant (eqs.18,24). The
results of [19] are shown by the crossed line for comparison.\\
  &\\
{\bf Fig.12} & The same as Fig.7 for the II variant (eqs.18,24).\\
  &\\
{\bf Fig.~13}  & Gluon distribution in the real photon, $Q^2=10 GeV^2$.
Thick solid line--our results (II variant), thin solid line -- our results
(I variant), dotted line -- the results of [18], dashed line -- the results
of [22], crossed line -- the results of [23]. \\
& \\
{\bf Fig.~14}  & Gluon
 distribution in the real photon; thick solid line -- our results, thin
 solid line, dotted and dashed lines -- the results of  fit [20].\\
 & \\
 {\bf Fig.~15} & Structure function $F_2$ of real photon for three
 flavours (thick solid line) at $Q^2_0=1 GeV^2$ for  $Q^2=5 \div 15 GeV^2$.
 Thin solid line and dotted line -- contribution from hadron part and gluon
 condensate, the later -- with opposite sign. Experimental  data
 (Refs.[24--29]) are shown for comparison.  Also results of
 Ref.[18,20] are shown by dashed and crossed lines respectively.
 (On Fig.15a results of PLUTO collab. are at $5.3 GeV^2$ and TPC - at $5.1
 GeV^2$.) \\
 & \\ {\bf  Fig.~16} & The same Fig.15, for $Q^2=20,45 GeV^2$. \\
 \end{tabular}

 \newpage
 \begin{tabular}{lp{12cm}}
 {\bf Fig.~17} & Structure function of transversally polarized virtual
 photon (thick solid line)  and unpolarized virtual photon (dashed line).
 Thin solid line and dotted line -- as in Fig.15.
  $p^2=-0.35 GeV^2,~Q^2_0=1 GeV^2$.\\
  & \\
  {\bf Fig.~18} & The same as Fig.17 for $p^2=-1 GeV^2,~-2 GeV^2$  and
  $Q^2_0=2 GeV^2,~4 GeV^2$ respectively. Cross line -- the results of
   Ref.[19].\\
   & \\
   {\bf Fig.~19} & Comparison of our results in
   $\overline{MS}$ (thin solid line) and $DIS_{\gamma}$ (thick solid line)
   schemes for gluon distribution and structure function in real photon
   case.  \\
   \end{tabular}


\newpage
\begin{thebibliography}{99}
\bibitem{1} A.S.Gorski, B.L.Ioffe, A.Yu.Khodjamirian, A.G.Oganesian Z.Phys.
{\bf C44} (1989) 523; ZhETF {\bf 97} (1990)  47.
\bibitem{2} B.L.Ioffe, Pisma ZhETF {\bf 42} (1985) 266, {\bf 43}
(1986) 316.
\bibitem{3} V.M.Belyaev, B.L.Ioffe, Nucl.Phys. {\bf B310} (1988) 548.
\bibitem{4} V.M.Budnev, I.F.Ginsburg, G.V.Meledin, V.G.Serbo, Phys.Pep.
{\bf 15} (1975) 181.
\bibitem{5} C.Peterson, C.H.Walsh, P.M.Zerwas, Nucl.Phys.
{\bf B174} (1980) 424.
\bibitem{6} H.Kolanovski -- Two photon
physics of $e^+e^-$  storage rings, -- Springer, Berlin, Heidelberg, New
York 1984.
\bibitem{7} Ch.Berger, W.Wagner Phys.Rep. {\bf 146} (1987) 1.
\bibitem{8} A.Levy, Report at International Workshop on Deep Inelastic
Scattering, Eilat, Israel, February, 1994.
\bibitem{9} V.A.Novikov, M.A.Shifman, A.I.Vanstein, V.I.Zakharov,
Ann.Phys. {\bf 105} (1977) 276.
\bibitem{10} E.Witten, Nucl.Phys. {\bf B120}(1977) 189.
\bibitem{11} M.Cl\"uck, E.Reya,   Phys.Rev~ {\bf D28} (1983) 2749.
\bibitem{12} W.A.Bardeen, A.J.Buras Phys.Rev. {\bf D20} (1979) 166;
{\bf D21}, 2041(E), 1980.
\bibitem{13} W.Furmanski, R.Petronzio Phys.Lett. {\bf 97B} (1980) 437.
\bibitem{14} J.De Witt, L.M.Jones, J.D.Sullivan, D.E.Willen and W.Wyld --
Phys.Rev.{\bf D19} (1979) 2046.
\bibitem{15} G.Rossi Phys.Rev. {\bf D29} (1984) 852.
\bibitem{16} M.Cl\"uck, E.Reya, A.Vogt Phys.Rev. {\bf D45} (1992)
3986.
\bibitem{17} T.Uematsu, T.F.Walsh  Nucl.Phys.{\bf B199} (1982) 93.
\bibitem{18} M.Gl\"uck, E.Reya, A.Vogt  Phys.Rev.{\bf D46} (1992)
1973.
\bibitem{19} M.Gl\"uck, E.Reya, M.Stratman, DO-TM 94/14.
\bibitem{20} K.Hagiwara, M.Tanaka, I.Watanabe, KEK-TH-376.
\bibitem{21} R.T.Herrod, W.Wada, B.R.Webber Z.Phys.{\bf C9} (1981) 351.
\bibitem{22} M.Dress, K.Grassie  Z.Phys. {\bf C28} (1985), 451.
\bibitem{23} H.Abramowicz, K.Grahula, A.Levy, Phys.Lett.~{\bf B269} (1991)
451.
\bibitem{24} $TPC/2\gamma$ Collab.,~ H.Aihara et al., Z.Phys.~ {\bf C34} (1987)
1.
\bibitem{25} $TPC/2\gamma$ Collab.,~ J.S.Sterman, UCLA preprint
UCLA-HEP-88-004 (1988).
\bibitem{26} PLUTO Collab.~ Ch.Berger et al., Z.Phys.~ {\bf C26} (1984) 353.
\bibitem{27} PLUTO Collab.,~ Ch.Berger et al. Phys.Lett. {\bf 142B} (1984)
111.
\bibitem{28} PLUTO Collab., Ch.Berger et al.~ Nucl.Phys. {\bf B281} (1987)
365.
\bibitem{29} OPAL Collab.,~ R.Akers et al.,~ Z.Phys. {\bf C61} (1994)
199-208.

\end{thebibliography}
\end{document}